\documentclass[5p,number]{elsarticle}

\usepackage{hyperref}
\usepackage[T1]{fontenc}
\usepackage[utf8]{inputenc}
\usepackage{lmodern} 
\usepackage{amsfonts}
\usepackage{breakurl} 
\usepackage[switch]{lineno} 
\usepackage{lipsum}
\usepackage{booktabs}
\usepackage{graphicx}
\usepackage{scrextend}
\usepackage{dcolumn}
\usepackage{bm}
\usepackage{textcomp}
\usepackage{mathtools}
\usepackage{float}
\usepackage{amsmath}
\usepackage{pifont}
\usepackage{natbib}
\usepackage{multirow}
\usepackage{mathrsfs}
\usepackage{amssymb}
\usepackage{geometry}
\usepackage{fleqn}
\usepackage{color}
\addtokomafont{labelinglabel}{\sffamily}
\biboptions{sort&compress}

\newcommand{\rco}{$R$Co$_2$}
\newcommand{\tbco}{TbCo$_2$}
\newcommand{\dyco}{DyCo$_2$}

\newcommand{\sioi}{Si (100)}

\newcommand{\tcl}{$T_{\rm C,\ Laves}$}
\newcommand{\dsm}{$\Delta S_{\rm M}$}

\newcommand{\dsmjj}{~mJ~cm$^{-3}$~K$^{-1}$}

\newcommand{\tcomp}{$T_{\rm comp}$}
\newcommand{\tbam}{Tb$_{31}$Co$_{69}$}
\newcommand{\dyam}{Dy$_{31}$Co$_{69}$}

\journal{Journal of Magnetism and Magnetic Materials}
\date{}
\begin{document}
\begin{sloppypar} 
\begin{frontmatter}

\title{Magnetocaloric effect in {\tbam} and {\dyam} thin films deposited on Si substrates}

\author[1]{P. Skokowski}\corref{cor1}
\ead{przemyslaw.skokowski@ifmpan.poznan.pl}
\cortext[cor1]{Corresponding author}
\author[1]{M. Matczak}
\author[1]{Ł. Frąckowiak}
\author[2]{T. Bednarchuk}
\author[1]{M. Kowacz}
\author[1,3]{B. Anastaziak}
\author[1]{K. Synoradzki}

\address[1]{Institute of Molecular Physics, Polish Academy of Sciences, Smoluchowskiego 17, 60-179 Pozna{\'n}, Poland}
\address[2]{Institute of Low Temperatures and Structural Research, Polish Academy of Sciences, Smoluchowskiego 17, 60-179 Pozna{\'n}, Poland}	
\address[3]{NanoBioMedical Centre, Adam Mickiewicz University, Wszechnicy Piastowskiej 3, 61-614 Poznan, Poland}

\begin{abstract}
We present the structural, magnetic, and magnetocaloric properties of thin films  with stoichiometry {\tbam} and {\dyam} deposited on naturally oxidized silicon {\sioi} substrates.
Samples with a thickness $d=50$~nm covered with a protective Au overlayer with a thickness $d_{\rm Au} =5 $~nm were produced using the pulsed laser deposition technique.
X-ray diffraction analysis indicated the presence of crystallized Laves phases and amorphous phases in the prepared materials.
Magnetization measurements as a function of temperature revealed ferrimagnetic behavior in both samples.
We estimated the compensation temperature {\tcomp} of the amorphous phase for {\tbam} at 81.5~K and for {\dyam} at 88.5~K, while we found the Curie temperature {\tcl} of the crystallized Laves phases at 204.5~K and at 117~K, respectively.
We investigated the magnetocaloric effect in a wide temperature range, covering {\tcomp} of amorphous phases and {\tcl} of crystallized Laves phases.
The analysis for the magnetic field change of $\Delta \mu_0H=5$~T showed values of the magnetic entropy change of $-\Delta S_{\rm M}=4.9$~{\dsmjj} at {\tcomp} and $-\Delta S_{\rm M}=6.6$~{\dsmjj} at {\tcl} for {\tbam}, while for {\dyam}, we determined the values of $-\Delta S_{\rm M}=35$~{\dsmjj} at {\tcomp} and $-\Delta S_{\rm M}=28$~{\dsmjj} at {\tcl}.
%
%

%
\end{abstract}


\end{frontmatter}

%
%
%
%
%
%
\section{\label{intr}Introduction}
%
%
%
%
%
%
Scientific investigation of materials involving rare-earth elements provided an opportunity to discover numerous physical phenomena, such as the Kondo effect, the heavy fermion state, unconventional superconductivity, or topological phases~\cite{li2019superconductivity, smidman2023colloquium, xu2023quantum}.
The realization of thin films of rare-earth alloys or multilayers utilizes the ferrimagnetic properties of the prepared materials.
One of the widely studied systems is Tb-Co, which may find application by involving all-optical switching, movement of domain walls, current-induced magnetization switching, or the creation of skyrmions~\cite{caretta2018fast, frkackowiak2020magnetic, aviles2020single, guo2021deterministic}.
Dy-Co systems have been studied in the context of magnetic memories in DyCo$_5$ or also skyrmions creation in DyCo$_3$ thin films~\cite{unal2016ferrimagnetic, chen2020observation, luo2023direct}.
The search for new applications for thin films containing rare-earth elements is especially promising for systems where various physical phenomena occur.
%

%
%

%
Magnetic materials containing rare-earth elements have been widely studied for magnetocaloric applications.
Although multiple promising bulk materials have been discovered, other forms of materials are being investigated.
Most known magnetocaloric materials, such as Gd, Gd$_5$(Si, Ge)$_4$, amorphous GdFeCo, or perovskites, have been verified in the form of thin films~\cite{kirby2013effects, miller2014magnetocaloric, hadimani2015gd5, matte2018tailoring, bouhani2020engineering, kumar2024broad}.
The results indicate that further studies may improve the magnetocaloric properties of some materials. 
%

%
%
One of the groups of intermetallic compounds with various magnetic properties is the Laves phase {\rco} ($R-$ rare earth element)~\cite{duc2002metamagnetism, singh2007itinerant, bonilla2010universal, balli2011magnetic}.
Within this group, the order of phase transitions changes from second-order type (SOT) for lighter rare-earth elements to first-order type (FOT) for heavier elements~\cite{khmelevskyi2000order}.
The phase transition type for {\tbco} is still discussed as FOT instead of SOT, as recent reports show~\cite{huang2020exotic, ahuja2015temperature, zhou2022unified}, while for {\dyco}, it is reported to exhibit FOT~\cite{morrison2013identifying}.
%
%
The {\rco} Laves phase is known for interesting properties connected with a magnetocaloric effect (MCE)~\cite{singh2007itinerant}.
Manipulation of the chemical composition of both compounds leads to modification of their magnetocaloric properties, for example, in TbCo$_{2-x}$Fe$_x$~\cite{jun2007magnetic, halder2010magnetocaloric, halder2011second}, Dy$_{1-x}$Er$_x$Co$_2$~\cite{cwik2011influence, cwik2013experimental}, and in many other systems based on {\rco} Laves phases, see references~\cite{gratz1995gd, ao2007magnetocaloric, zhou2006magnetocaloric, burrola2012magnetocaloric, cwik2012influence, zheng2018magnetocaloric, wang2019magnetic, murtaza2020magnetocaloric}.
These two compounds have also been prepared in non-bulk forms, exemplary, such as {\tbco} as thin films in the form of multilayer~\cite{robaut1996epitaxial, jaren1997pulsed, grenier2007structural, ovcharenko2022ultrafast}, or {\dyco} as nanoparticles~\cite{ma2008large}.
The amorphous {\rco} phases were realized in the hydrogenation process~\cite{mushnikov2005magnetic}.

In this article, we present magnetic and magnetocaloric studies of thin films with stoichiometry {\tbam} and {\dyam} deposited using the pulsed laser deposition (PLD) technique.
The thickness of the thin films studied was $d=50$~nm for both materials.
The films were prepared on naturally oxidized silicon {\sioi} substrates.
X-ray diffraction (XRD) measurements revealed the presence of crystallized {\rco} Laves phases and amorphous phases.
The magnetic properties of samples were investigated in the form of magnetization measurements as a function of temperature and magnetic field.
For both samples, the analysis of the first derivative revealed compensation temperatures {\tcomp} below 100~K of the amorphous phase and Curie temperatures {\tcl} of the crystallized {\rco} Laves phases.
As the magnetocaloric effect of the {\tbco} and {\dyco} Laves phase presents promising magnetocaloric parameters, the samples were studied in a wide range of temperatures to cover {\tcomp} of the amorphous phases and also {\tcl} of the corresponding bulk compounds~\cite{singh2007itinerant}.
On this basis, the magnetocaloric parameters were determined: magnetic entropy change {\dsm} and temperature-averaged entropy change $(TEC)$.
\begin{figure*}[h!]
\centering
\includegraphics[width = 1.8\columnwidth]{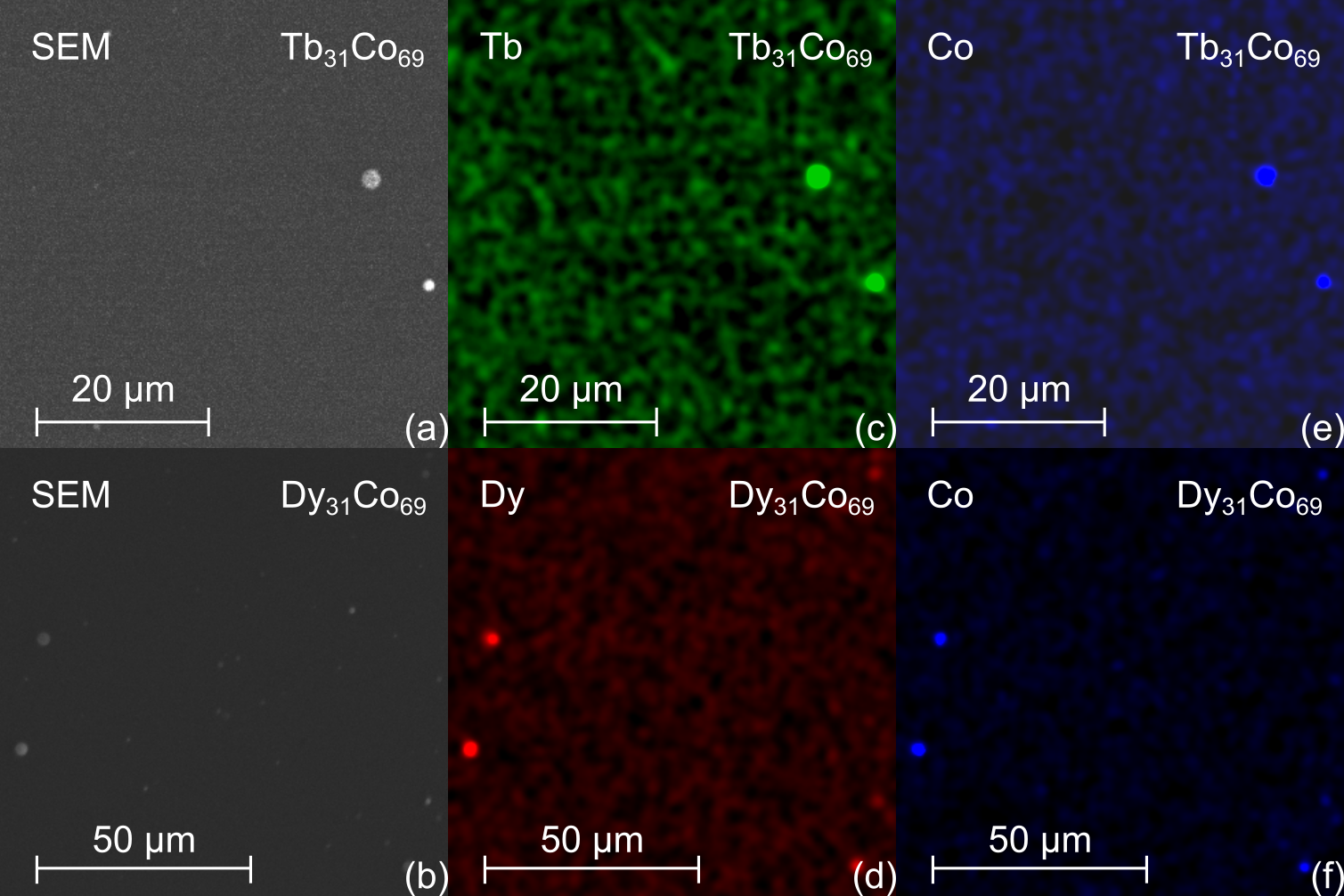}
\caption{\label{sem1}%
(a) and (b) Topography of the {\tbam} and {\dyam} layers. (c-f) Chemical maps of specific elements for the samples with the {\tbam}  and {\dyam} layers.
}
\end{figure*}
%
%
%
%

\section{\label{exp}Experimental}
%
%
%
%
%
The layers were deposited from commercially produced targets (Kurt Lesker) with a nominal stoichiometry of 1:2 of Tb and Dy to Co using the PLD technique (Nd:YAG laser).
The laser wavelength was $\lambda = 532$~nm with a frequency of 2~Hz.
The chamber pressure was about $8\times10^{-8}$~mbar.
Thin films were deposited on naturally oxidized {\sioi} substrates.
The nominal thickness of the Tb-Co and Dy-Co layers was $d=50$~nm.
The layers were covered with an Au layer $d_{\rm Au}=5$~nm deposited using the same PLD parameters to protect against oxidation.
The stoichiometry of the prepared samples and their topography were studied by scanning electron microscopy (SEM) and energy dispersive spectroscopy (EDS).
The device used for this purpose was a~FEI Nova NanoSEM 650 equipped with an EDS Bruker detector.
The crystal structure of the prepared samples was examined by XRD measurements at room temperature using the X’pert Pro PANalytical device with a Cu $K \alpha$ radiation source.
Magnetic properties investigation was performed on the Quantum Design Physical Property Measurement System (QD PPMS) in DC magnetization mode using a vibrating sample magnetometer (VSM).
Measurements of magnetization were collected as a function of temperature in zero-field cooling (ZFC) and field cooling (FC) modes.
The experiments were carried out in the temperature range of 10-380~K with a magnetic field of 0.1~T.
The isothermal magnetization was measured as a function of the applied magnetic field at magnetic field values up to 5~T in various temperature ranges.
%


%

%

%

%
%
%
%

%

%
\section{\label{res}Results}
%
%
%
%
%
%
%
\subsection{\label{eds}Microstructure investigation}
The first step was the verification of the chemical composition of the samples using EDS.
The atomic percentage results showed that the stoichiometry of all samples was close to the nominal stoichiometry.
For the Tb-Co layer, the atomic percentage content was 30.8(1.4)\% of Tb and 69.2(1.4)\% of Co, while for the Dy-Co layer, it was 30.9(0.7)\% of Dy and 69.1(0.7)\% of Co.
The nominal atomic percentage content is expected to be 33\% of Tb or Dy and 67\% of Co.
Due to stoichiometry, the samples are labeled {\tbam} and {\dyam}.
For both samples, the SEM images presented droplets of material on the surface of the layers, which is characteristic of the PLD method, see Figures~\ref{sem1}(a) and (b).
Chemical maps shown in Figures~\ref{sem1}(c-f) indicated the uniform distribution of all elements on the sample surface.

\subsection{\label{xrdsr}X-ray diffraction}
\begin{figure*}[h!]
\centering
\includegraphics[width = 2.0\columnwidth]{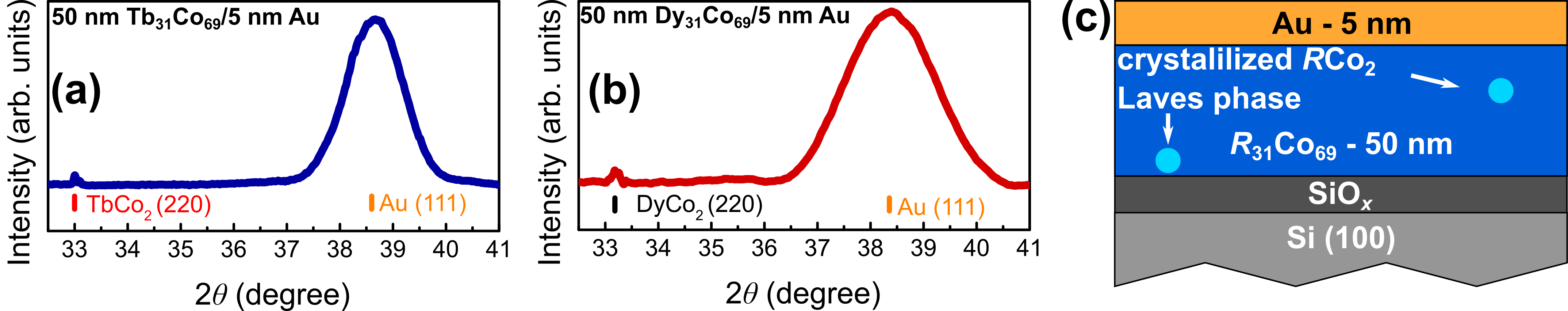}
\caption{\label{xrd}%
(a) and (b) X-ray diffraction patterns of the {\tbam} and {\dyam} layers.
Wide maxima at $2\theta = 38.3^{\circ}$ are related to the Au (111) peak, small maxima at $2\theta = 33.1^{\circ}$ are related to the {\tbco} and {\dyco} (220) peaks.
%
%
(c) Schematic illustration of the samples studied.
Small light blue balls indicate crystals of the {\rco} Laves phase.
}
\end{figure*}
%
The XRD results for the samples with the {\tbam} and {\dyam} layers are shown in Figures~\ref{xrd}(a) and (b).
The results for the entire $2\theta$ range for both samples (not shown) revealed two major maxima, one at $2\theta =69.3^{\circ}$, which is related to the Si (400) peak, and the other allocated at $2\theta = 38.3^{\circ}$, which is related to the Au (111) peak.
For both samples, in the vicinity of the Au (111) peak, small peaks are observed around $2\theta = 33.1^{\circ}$.
These reflections were assigned to a close approximation of the position of the (220)peak of the MgCu$_2$-type structure of {\tbco} and {\dyco} Laves phases~\cite{khmelevskyi2000order, chang2019crystal}.
The peak position is comparable for both compounds, suggesting that the values of the lattice parameters $a$ are very similar, although slight differences in the crystal lattice parameters of bulk {\tbco} and {\dyco} were observed~\cite{khmelevskyi2000order, chang2019crystal}.
Although the presence of Laves phases is noted, the low intensities of the peaks of the (220) peaks suggest that the crystallized volume is small.
However, significant widening of the Au (111) peaks may suggest that these protective layers grew mainly on the amorphous {\tbam} and {\dyam} layers, even if their amorphous halos are not visible. 
Therefore, the results should be interpreted as evidence of the presence of crystallized Laves phases in the majority of the amorphous layers.
The schematic composition of the samples is presented in Figure~\ref{xrd}(c).
\begin{figure*}[h!]
\centering
\includegraphics[width = 2.0\columnwidth]{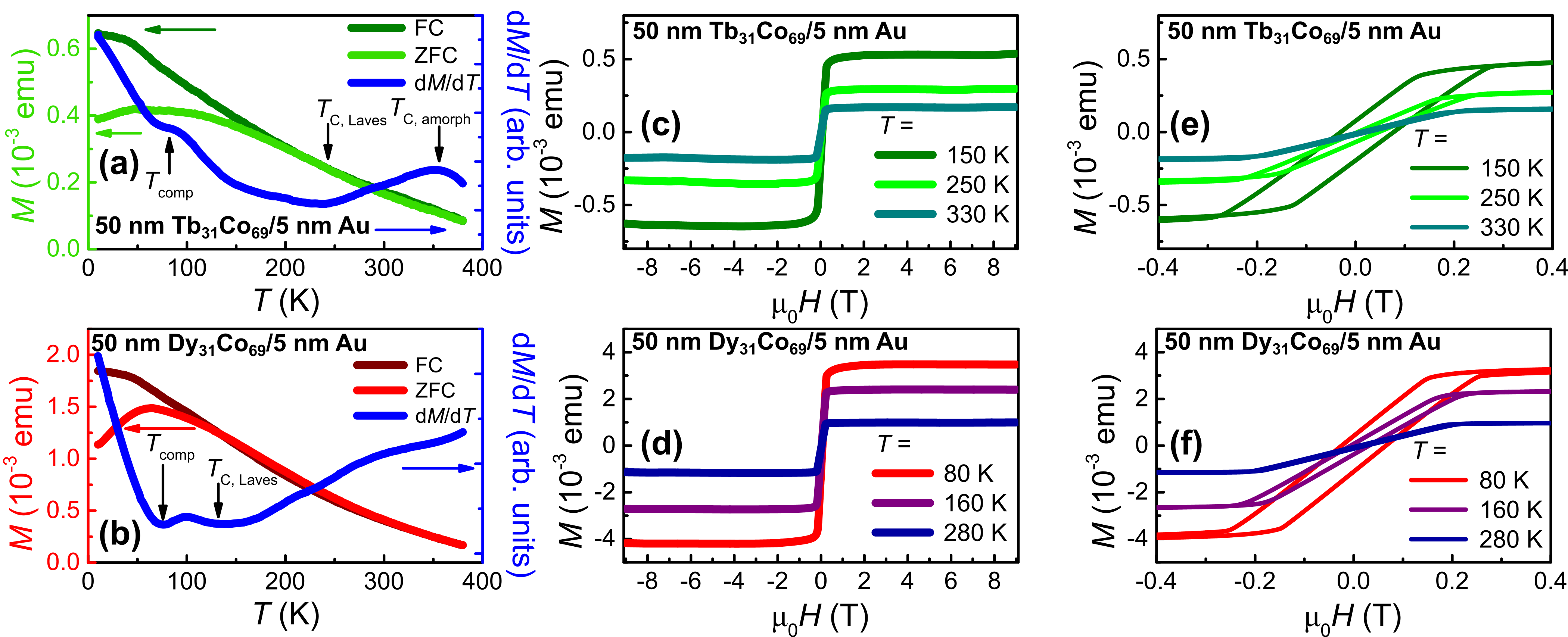}
\caption{\label{magnetic}%
Magnetic properties of the samples with the {\tbam} and {\dyam} layers.
(a) and (b) Magnetization as a function of temperature measured in a magnetic field value of 0.1~T for for the {\tbam} and {\dyam} layers. 
The right blue axis is assigned to the derivative of the FC magnetization. 
%
%
The amorphous phase compensation temperatures are denoted as {\tcomp}.
The Curie temperatures of crystallized Laves phases are denoted as {\tcl}.
For {\tbam} at around 350~K $T_{\rm C,\ amorph}$ of the amorphous phase is noted.
(c) and (d) Magnetization isotherms as a function of an applied magnetic field for the {\tbam} and {\dyam} layers.
(e) and (f) Close up on the hysteresis of the magnetization isotherms as a function of an applied magnetic field for the {\tbam} and {\dyam} layers.
}
\end{figure*}

\subsection{\label{msus}Magnetic properties}
%

%
%

%
Magnetization as a function of temperature is presented in Figure~\ref{magnetic}(a) and (b) for samples with {\tbam} and {\dyam} layers.
The ZFC and FC curves show no distinct magnetic phase transitions for both samples.
The main visible feature of the results obtained is a bifurcation of these curves, observed at 200~K for the sample with the {\tbam} layer and at 120~K for the sample with the {\dyam} layer.
The first derivative was used to verify the presence of phase transitions.
Several local extrema are registered for both samples.
The first local extrema are detectable in the temperature range of 100~K for both layers.
%
%
This observation is in good approximation with the results for the amorphous {\rco} obtained in the hydrogenation process, with $T_{\rm comp}=98$~K for the {\tbco} stoichiometry and $T_{\rm comp}=93$~K for the {\dyco} stoichiometry~\cite{mushnikov2005magnetic}.
The second local extrema are observed around 230~K and 130~K for the samples with {\tbam} and {\dyam} layers, respectively.
These temperatures correspond to {\tcl} of the {\tbco} ($T_{\rm C,\ bulk}=230$~K) and {\dyco} ($T_{\rm C,\ bulk}=135$~K) Laves phases.
However, in the ZFC and FC curves, the transitions are barely visible as an inflection point, which confirms the minimal contribution of the crystallized Laves phases.
Furthermore, for the sample with the {\tbam} layer, a local maximum of the derivative around 350~K was registered.
The maximum is related to $T_{\rm C,\ amorph}$ of the amorphous phase~\cite{soltani2001composition}.
Notably, the magnetization results also confirm the presence of crystallized Laves phases in the majority of the amorphous layers indicated in the XRD analysis.
Magnetization as a function of the magnetic field was examined at different temperatures for both samples.
The results are presented in Figures~\ref{magnetic}(c) and (d).
The magnetization curves show ferrimagnetic behavior for all tested temperatures for both layers.
The magnetization values increase with decreasing temperature.
The close-up of the hysteresis is shown in Figures~\ref{magnetic}(e) and (f).
In both cases, the hysteresis is not noticeable at the highest measured temperatures.
The hysteresis increases with decreasing temperatures with coercivity up to 0.1~T at 150~K for the {\tbam} layer and up to 0.09~T at 80~K for the {\dyam} layer.

\subsection{\label{mce}Magnetocaloric properties}

\begin{figure*}[t!]
\centering
\includegraphics[width = 2.0\columnwidth]{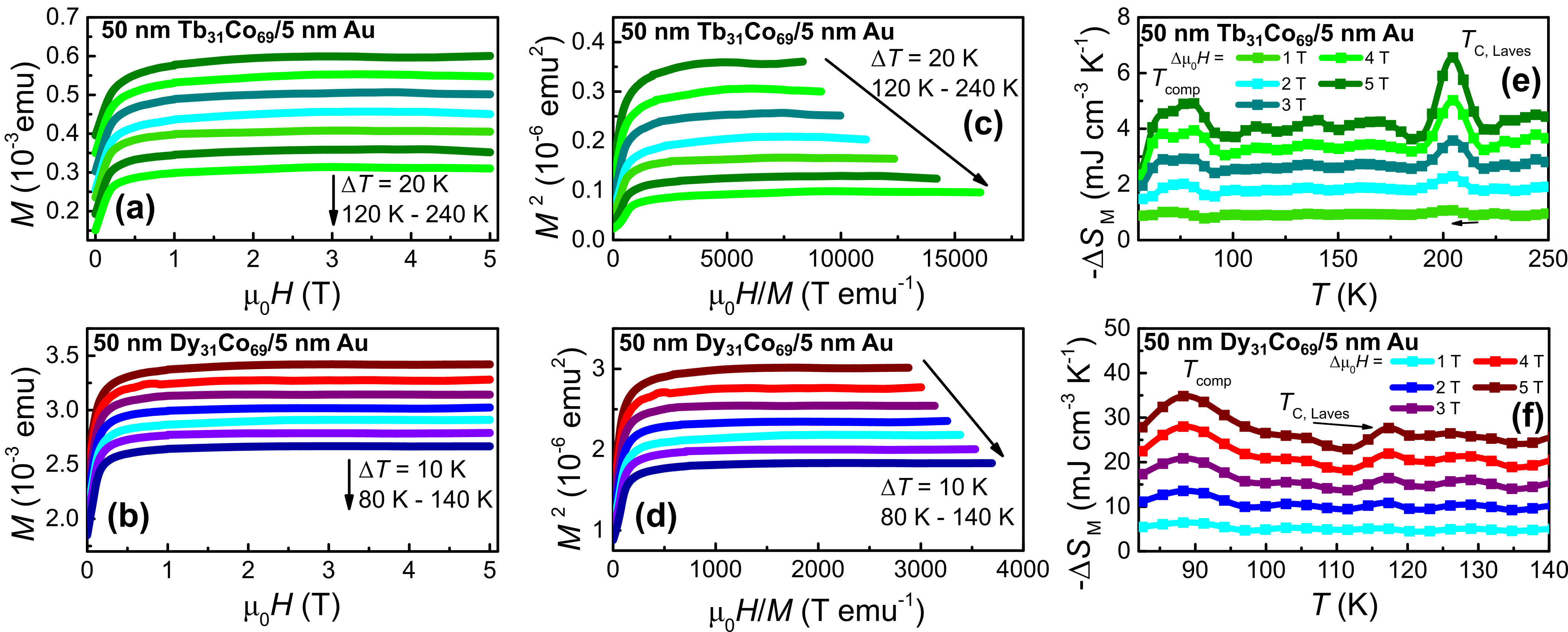}
\caption{\label{mce}%
%
%
(a) and (b) Magnetization isotherms as a function of magnetic field measured at exemplary temperatures for the samples with {\tbam} and {\dyam} layers.
(c) and (d) Arrott plots for the samples with {\tbam} and {\dyam} layers.
(e) and (f) Magnetic entropy change {\dsm} for the {\tbam} and {\dyam} layers for various magnetic field values.
}
\end{figure*}


%

The magnetocaloric effect was determined for the samples studied in the temperature range, which includes {\tcomp} of the amorphous majority of the samples and {\tcl} of the small contribution of the Laves phases.
%
%
The first quarters of the magnetization isotherms as a function of the magnetic field collected at exemplary temperatures for samples with {\tbam} and {\dyam} layers are presented in Figure~\ref{mce}(a) and (b).
Figure~\ref{mce}(c) and (d) show Arrott plots in the form of $M^2$ as a function of $\mu_0H/M$.
The samples do not exhibit negative curvature, indicating the second-order type phase transition~\cite{arrott1957criterion}.
For bulk {\tbco} and {\dyco} Laves phases, the first-order type phase transition is expected~\cite{huang2020exotic, ahuja2015temperature, zhou2022unified, morrison2013identifying}.
After collecting the experimental results of the magnetization isothermal measurements, the magnetic entropy change {\dsm} could be calculated according to the formula~\cite{gschneidner2005recent}:
\begin{equation}
\begin{aligned}
\Delta S_{\rm M} &\approx \frac{\mu_{0}}{\Delta T}\Big[\int_0^{H_{\max}} M(T+\Delta T, H)\mathrm{d}H \\
&- \int_0^{H_{\max}}M(T,H)\mathrm{d}H\Big],
\end{aligned}
\end{equation}
where: $\mu_0$ -- the magnetic permeability of the vacuum, $H_{\max}$ -- the maximum magnetic field value to calculate the specific $\Delta S_{\rm M}(T, H_{\max})$, $\Delta T$ -- the temperature difference between two closest magnetization isotherms, $M(T, H)$ -- magnetization for the specific magnetic field value and temperature $T$, $M(T + \Delta T, H)$ -- magnetization for specific magnetic field value and temperature $T + \Delta T$.
%
To describe how the {\dsm} peak spans over a temperature range, the temperature-averaged entropy change $TEC$ was calculated for lift temperatures of $\Delta T_{\rm lift}=3$~K and 10~K.
The value of this parameter is calculated using the following formula~\cite{griffith2018material}:
\begin{flalign}
TEC(\Delta T_{\rm lift}) =\frac{1}{\Delta T_{\rm lift}} \max\displaylimits_{T_{\rm mid}} \left\{ \int\displaylimits_{T_{\rm mid}-\frac{\Delta T_{\rm lift}}{2}}^{T_{\rm mid}+\frac{\Delta T_{\rm lift}}{2}}\Delta S_{\rm M}(T)_{\mu_0H}\mathrm{d}T \right\},
{\label{eqTEC}}
\end{flalign}
where: $\Delta T_{\rm lift}$ -- the desired lift temperature, $T_{\rm mid}$ -- the temperature of the center of the TEC calculated to maximize the TEC value.
Another magnetocaloric parameter, the relative cooling power $RCP$, was not determined because, in the samples studied, the full width at half the maxima of the peaks could not be found in the measured temperature range.

%
%
In order to evaluate the {\dsm} values in {\dsmjj} units for the deposited $R$-Co layers, it was necessary to determine the volume of the layers.
The volume of the deposited layers was assumed to be the area of the samples multiplied by the thickness of the layers.
The {\dsm} results obtained for the deposited $R$-Co layers are shown in Figures~\ref{mce}(e) and (f), while the values of all magnetocaloric parameters are gathered in Table~\ref{tab3}.
For both samples, two maxima connected to the amorphous {\tcomp} and to the {\tcl} of the Laves phases are observed.
In the case of the {\tbam} sample, the {\tbco} Laves phase-related peak is more pronounced and is at 204.5~K, shifted towards lower temperatures compared to the bulk crystallized {\tbco} sample.
The amorphous {\tcomp}-related maximum is at 81.5~K.
The maximum values for both maxima are very low, below 7~{\dsmjj} for the magnetic field change of $\mu_0H= 5$~T, see Table~\ref{tab3}.
%
%
The width of the {\dsm} maxima described by $TEC$ indicates that for both peaks, the {\dsm} values do not change significantly, even for $TEC(10)$.
For the {\dyam} layer, the {\dsm} maximum related to the amorphous {\tcomp} is at a temperature 88.5~K, and another less pronounced maximum is probably related to the {\dyco} Laves phase at 117~K.
The values of the obtained {\dsm} at {\tcomp} is 35~{\dsmjj}, while at {\tcl} of the Laves phase is 28~{\dsmjj}.
%
%
%
$TEC(3)$ and $TEC(10)$ values for this sample also indicate small changes of {\dsm} over the specific temperature range.
%

%

%

%
%
It is important to note that both samples are in a ferrimagnetic state in the temperature range studied, since the $T_{\rm C,\ amorph}$ of the amorphous {\rco} alloys is over 300~K~\cite{takahashi1989perpendicular, soltani2001composition}.
Therefore, the values of magnetocaloric parameters are relatively low.
However, the contribution to {\dsm} of {\tcomp} is noticeable even in the ferrimagnetic state.
While for the sample with {\dyam} layer, {\dsm} has higher values at {\tcomp}, for the sample with {\tbam} layer, {\dsm} at {\tcl} of the crystallized Laves phase has higher values.
Although the Laves phase contribution is observed, the content of this phase in the sample volume is low, resulting in a low contribution to the {\dsm} values.
%
%

\begin{table*}[t!]
\caption{\label{tab3}
Maximum value of magnetic entropy change {\dsm}$^{\max}$, temperature of maximum {\dsm}$^{\max}$ value $T_{max}$ and temperature averaged entropy change $TEC(3)$ and $TEC(10)$ for the {\tbam} and {\dyam} layers.
The parameter values were obtained for the magnetic field change of $\Delta \mu_0H=5$~T.
For both samples the determined values refer to maxima related to the {\tcomp} of amorphous phase and to the {\tcl} of the crystallized Laves phase.
%
%
}
\centering
\def\arraystretch{1.5}%
\begin{tabular}{ccccc}
\hline
Parameters & {\tbam}, {\tcomp} & {\tbam}, {\tcl} & {\dyam}, {\tcomp} & {\dyam}, {\tcl}\\
\hline
{\dsm}$^{\max}$ (mJ~cm$^{-3}$~K$^{-1}$) & 4.9  & 6.6 & 35 & 28\\
$T_{max}$ (K) & 81.5 & 204.5 & 88.5 & 117\\
$TEC(3)$ (mJ~cm$^{-3}$~K$^{-1}$) & 4.8 & 6.4 & 34 & 27\\
$TEC(10)$ (mJ~cm$^{-3}$~K$^{-1}$) & 4.7 & 6.2 & 33 & 26\\
\hline
\end{tabular}
\end{table*}

\section{\label{sum}Summary}
Thin films of {\tbam} and {\dyam} were prepared and deposited on a naturally oxidized Si (100) substrate.
The thickness of the layers was $d = 50$~nm covered with an Au overlayer with a thickness of $d_{\rm Au} = 5$~nm.
Scanning electron microscopy and energy dispersive spectroscopy showed a homogeneous chemical distribution with atomic stoichiometry of {\tbam} and {\dyam}.
The presence of crystallized Laves phases in the layers was verified by X-ray diffraction, which revealed low-intensity maxima associated with the (220) peak for both thin films.
The widening of the Au (111) peaks indicated that the {\tbam} and {\dyam} layers are mainly amorphous.
Analysis of the magnetization results as a function of temperature for both samples revealed the compensation temperature of the amorphous phases and the Curie temperature of the corresponding Laves phase compounds.
The magnetization as a function of the magnetic field indicated ferrimagnetism for both samples at all measured temperatures.
The magnetocaloric effect was investigated in a wide temperature range.
The obtained results of the magnetic entropy change {\dsm} revealed maxima connected with the compensation temperature {\tcomp} of the amorphous phase and with phase transitions connected with the presence of crystallized Laves phases.
These contributions to {\dsm} are visible in the ferrimagnetic state of the {\tbam} and {\dyam} layers.
\section*{Acknowledgments}

PS acknowledges the financial support of the National Science Centre Poland under the decision 2021/05/X/ST5/ 00763.
The authors would like to thank Piotr Kuświk from the Department of Thin Films and Nanostructures IMP PAS for the discussion and helpful comments.

\bibliography{bib}

\end{sloppypar}
\end{document}